\documentclass[twocolumn,english]{revtex4}
\usepackage[T1]{fontenc}
\usepackage[latin9]{inputenc}
\usepackage{amsmath}
\usepackage{graphicx}

\makeatletter
\@ifundefined{textcolor}{}
{%
 \definecolor{BLACK}{gray}{0}
 \definecolor{WHITE}{gray}{1}
 \definecolor{RED}{rgb}{1,0,0}
 \definecolor{GREEN}{rgb}{0,1,0}
 \definecolor{BLUE}{rgb}{0,0,1}
 \definecolor{CYAN}{cmyk}{1,0,0,0}
 \definecolor{MAGENTA}{cmyk}{0,1,0,0}
 \definecolor{YELLOW}{cmyk}{0,0,1,0}
 }

\makeatother

\usepackage{babel}

\begin{document}

\title{Nonequilibrium phase transitions in active contractile polar filaments}

\author{Kripa Gowrishankar$^{*}$ and Madan Rao$^{*,**}$}

\affiliation{$^{*}$Raman Research Institute, C.V. Raman Avenue, Sadashivanagar,
Bangalore 560080, India\\
 $^{**}$National Centre for Biological Sciences (TIFR), Bellary Road, Bangalore 560065,
India}
\begin{abstract}
We study the patterning and fluctuations of a collection of active contractile polar filaments on a two dimensional substrate, using a continuum description in the presence of athermal noise, parametrized by an active temperature $T_A$. The steady states generically consist of arrays of inward pointing asters and show a continuous transition from a moving aster street to a stationary aster lattice. In contrast to its equilibrium counterpart,  this active crystal shows true long range order
at low $T_A$.
On increasing $T_A$, the asters remodel with a distribution of lifetimes; concomitantly we find novel phase transitions characterized by polar and bond-orientational order.
\end{abstract}

\maketitle

\begin{figure}[t]
\includegraphics[scale=0.8]{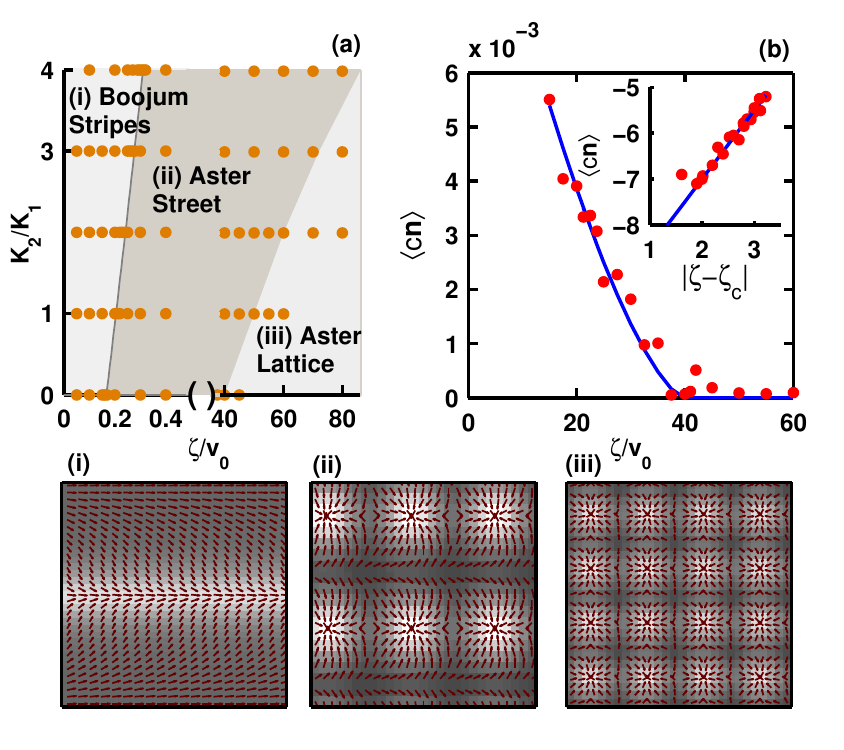}
\caption{(a)  Phase diagram in $K_2/K_1$ versus $\zeta/v_0$ at $T_A=0$, where $D=2.5$ and
  $\alpha=\beta=100$ in dimensionless units (see text), shows three phases - (i) Boojum stripes (ii) Aster street and (iii) Aster lattice.
Representative configurations of $c$ (shading) and $\bf{n}$ (arrows), when $K_1=2.5$, $K_2=0$, $v_0=1$,
  and $\zeta= 0.1\, \mbox{(i)}, 5\, \mbox{(ii) and} \,80 \,\mbox{(iii)}$ are shown alongside.
  (i) and (ii) are {\it moving} phases, while (iii) is a stationary phase.   
  (b)  The net polarization $\langle c\bf{n}\rangle$ goes to zero continuously, as the contractility parameter $\zeta/v_0$ changes across the moving to stationary phases, with a power law behaviour
 $\langle c\bf{n}\rangle \sim \vert\zeta-\zeta_c\vert^\gamma$, where $\zeta_c=40$ and $\gamma\approx1.4$ ({\it Inset} shows a logarithmic plot (base $e$),
 lines show fit).}
\end{figure}

A remarkable feature of living cellular systems is  that the same evolutionarily conserved ingredients -- filaments (actin), motors (myosin) and their regulators, in the presence of ATP --  can exhibit a variety of phenotypes depending upon the cellular context \cite{review}. Recently, there have been serious 
attempts to quantitatively understand this spectrum of behaviours using in-vitro reconstituted systems \cite{bausch, shamim}.  
In this paper, we
provide a detailed analysis of 
the nonequilibrium steady states and phase transitions of actively driven collections of filaments-motors in two dimensions using the  framework of {\it active hydrodynamics}
or {\it active gels} \cite{sriramreview,joannyreview}, which could in principle be used to compare
with such experiments.

Our study differs from earlier studies of active hydrodynamics in two  important aspects : (i) We provide an exhaustive treatment of  the steady states and their transitions 
using dynamical equations written in terms of {\it both} the concentration and polarization of the active filaments. Realizing that there are many different
microscopic processes, such as acto-myosin contractility and treadmilling, that may simultaneously engage with the  cortical actin, has allowed us to probe parameter regimes that have not been explored before.
 (ii) We study the effects of spatiotemporal active noise that are inevitably present in cellular systems. This not only affects the
dynamics in the steady state, but also induces novel phase transitions that are characterized by a variety of order parameters. Our method of analysis combines both analytical (linear stability analysis) and detailed numerical solutions and complements \cite{shraddha}. Our main results: (a) We find a variety of steady state 
configurations that include domain walls, boojums, inward-pointing asters and spirals; 
vortices are generically unstable. The phase diagram, which includes a transition from a moving aster street phase to a stationary aster lattice phase across a critical 
line (Fig.\,1), is robust and is our first main result. (b) The 2d active aster lattice phase is stable even in the presence of noise; it shows true long-range order (LRO), in striking contrast to its equilibrium counterpart (Fig.\,2).
(c) Beyond a critical active temperature, this transforms to an aster lattice with quasi long range order (QLRO, Fig.\,2) during which the asters remodel with a power law distribution of lifetimes. This phase exhibits strong bond tetratic order (Fig.\,3). (d)
On increasing the active temperature further, we observe a discontinuous transition to a bond nematic phase (Fig.\,3) with an exponential aster lifetime distribution. Significantly, this bond nematic phase is polar, the filaments have a net orientation and exhibit power-law orientational correlations in this phase (Fig.\,4), with exponents in the Toner-Tu universality class \cite{toner-tu}.
 
We describe the acto-myosin system as a collection of permanent force-dipoles, 
which being polar, will induce a net drift of the $i^{th}$ filament with respect to the medium, represented by a polarization vector ${\bf n}_i$. 
The hydrodynamic fields are the local concentration $c({\bf r},t)$, the polarization density $c({\bf r},t){\bf n}({\bf r},t)$, and the hydrodynamic velocity ${\bf v}({\bf r},t)$ \cite{simha-sriram,curie1,marchetti1}.  

We will assume that momentum is lost by local friction at the ``substrate'', thus $\Gamma {\bf v}=-\nabla \cdot \sigma$, where $\Gamma$ is the friction coefficient and $\sigma \propto c {\bf n} {\bf n}$ is the active stress \cite{hatwalne} due to the force-dipoles.
We can use this to eliminate ${\bf v}$ from the equations of ${\bf n}$ and $c$.

The hydrodynamic equations for active filaments (undergoing contractility and treadmilling) can be written as \cite{toner-tu},
\begin{eqnarray}
\partial_{t}{\bf n}+{\lambda({\bf n}\cdot\nabla){\bf n}} & = & {K_1\nabla^{2}{\bf n}+K_2\nabla(\nabla\cdot{\bf n})}+{\zeta\nabla c}\nonumber \\
\,\,\,\,\, &  & {+\alpha{\bf n}-\beta\vert{\bf n}\vert^{2} {\bf n}}+{\bf f} \label{nequation}\\
\partial_{t}c=-\nabla\cdot{\bf J} & = & -\nabla\cdot\left({v_{0}\, c\,{\bf n}}-D \nabla c\right)\label{cequation}\end{eqnarray}
 to  lowest order in gradients and fields (contributions from ${\bf v}$ appear at  higher order).
 The right hand side of (\ref{nequation}) and (\ref{cequation})
represent contributions to active forces/torques and current ${\bf J}$.
The parameter $\alpha$ measures the deviation of the mean filament concentration ${\bar c}$ from the Onsager value which fixes the transition to orientational order.
We choose a value of $\alpha$ and $\beta$, such that the magnitude of ${\bf n}$ in the ordered phase is close to unity in most places.
The terms $v_0$ and $\lambda$ are uniquely active in origin, and represent an active advection and a nonlinear active convective contribution, respectively \cite{note-convective}.
The athermal noise ${\bf f}$ is taken to be white with zero mean and variance equal to $T_A/c$, where $T_A$ is the active temperature. We have dropped the additive noise term in the $c$ equation, since the multiplicative nonequilibrium driving from the first term in ${\bf J}$ is more dominant.

The values of these parameters depend on the microscopic active processes controlling actin dynamics. In a typical cellular context there are several microscopic active processes occurring simultaneously, such as treadmilling and actomyosin contractility.
We therefore independently vary the parameters entering the dynamical equations over a  range of values. Note that for the contractile motor-filament system, $\zeta > 0$,  which is {\it opposite} to the flocking case considered in \cite{toner-tu}. 
With these parameters, one can construct the following independent  length scales -- (i) correlation lengths $L_c$, given by $\sqrt{K_1/\alpha}$ and 
 $\sqrt{K_2/\alpha}$, (ii) extrapolation lengths $L_e$, given by $K_1/\zeta$ and $K_2/\zeta$ and (iii) Peclet length, $L_p = D/v_0$, the ratio of diffusion coefficient to advection. We will predominantly work in the regime where  both $L_c$ and $L_e$ are small, further, we study the phase diagram when the magnitude of $\lambda$ is zero or small.
 
We first explore the phase diagram when the active temperature, $T_A=0$. We find that the uniform orientationally disordered phase is stable when $\alpha<0$ and $\zeta>0$ but small, but beyond a threshold 
$\zeta =-D \alpha/v_0$, there is a clumping instability whose scale is set by the inverse of the maximal unstable
mode $k_{d} =  \sqrt{\delta/K_1D}$, where $\delta=\zeta v_0+D\alpha$.
On the other hand, the uniform orientationally ordered phase which is stable when $\alpha>0$ and $\zeta=0$, is spontaneously unstable 
to splay distortions as soon as $\zeta v_0 > 0$. Taking the ordering direction to be along $\hat{\bf x}$, we find that there is a band of unstable 
wavevectors centered around $k_{x}=0,k_{y}\equiv k_0 =  \frac{\sqrt{\zeta v_{0}}}{(D+K_1+K_2)}$.

\begin{figure}[t]
\begin{centering}
\includegraphics[angle=0,scale=0.6]{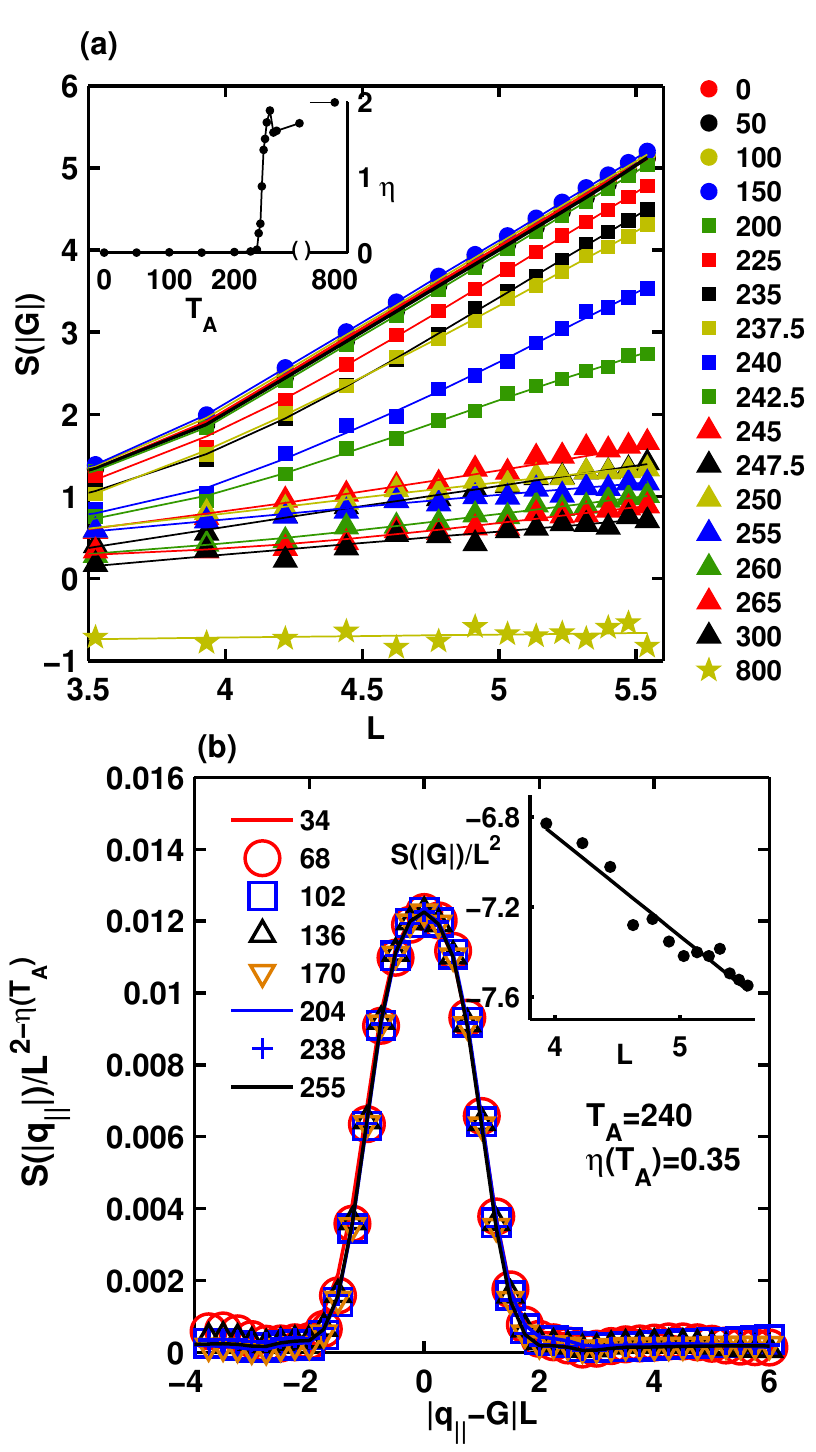}
\par\end{centering}
\caption{(a) Logarithmic plot (base $e$) of the peak amplitude of the aster-density structure factor $S$ evaluated at the reciprocal lattice vector ${\bf G}$ of the square crystal, as a function of 
system size $L$, over a range of active temperatures $T_A$ (displayed symbols to the right). For $0<T_A<160$, this amplitude scales as $L^2$, suggesting true long range
positional order. Beyond this temperature, the amplitude scales as $L^{2-\eta}$, where the critical exponent $\eta$ is a function of $T_A$ (inset), suggesting a transition to a phase with QLRO. At still higher active temperatures, there is a transition to a short range order (SRO).
(b) At $T_A=240$, corresponding to the QLRO phase, the structure factor evaluated about the peak ($q_{\|}$ is parallel to ${\bf G}$) exhibits a finite-size scaling form. 
The different system sizes $L$ are shown as symbols. Inset shows a logarithmic plot (base $e$) of the scaling of the peak amplitude of $S$ with $L$ with a value of $\eta$.
 Here, $\zeta= 100$ and the rest of the parameters as in Fig.\,1.
}
\end{figure}

The final steady state configurations when $\alpha$ and $\zeta$ are positive, depend on the extrapolation length $L_e$ relative to the other lengths, and can be obtained by numerically solving
Eqs.(\ref{nequation}),(\ref{cequation}). For this purpose, it is convenient to convert the equations to dimensionless form, by choosing the units of
length, time and $\vert {\bf n}\vert$ to be $D/2.5v_0$, $D/2.5v_0^2$ and $\sqrt{\alpha/\beta}$, respectively. The values of the various parameters chosen for the numerics
are written in these units.
We use an implicit {\it alternate direction operator splitting} scheme with length and time discretizations chosen to be $\Delta x =1$ and $\Delta t=0.01$ respectively \cite{numrecipe}, with initial conditions for  $c(\bf{r},t)$ and $\bf{n}(\bf{r},t)$ being homogeneous and random. The boundary conditions are chosen to be periodic; however, our results hold for other boundary conditions as well, as long as the system size $L \gg L_c$ (when boundary effects negligible). In order 
to maintain conservation and non-negativity of the local concentration, we use symmetric spatial derivatives (that add up to zero over the whole system) and choose an adaptive (small) grid size $\Delta t$.

%

Since $\zeta>0$, the steady state configurations generically consist of a collection of 
defects such as Boojums, inward-pointing asters \cite{leekardar,sumithra}, inward-pointing spirals or walls. This might be expected, since the filament current ${\bf J} \propto {\bf n}$ and the steady state equations for ${\bf n}$ is roughly a vector Poisson equation with $\frac{\zeta}{K_1} \nabla c$ as source.


We have explored the small extrapolation length regime in some detail, where
upon increasing $\zeta$, we encounter the following defect phases (Fig.\,1) : 
(i) {\it Boojum stripes} - consisting of alternate stripes of filaments oriented along ${\hat {\bf x}}$ and  configurations associated with a $+2$ defect known as ``boojum'' configurations with scale $1/k_{0}$. (ii) {\it Aster street} - consisting of alternate stripes of filaments oriented along ${\hat {\bf x}}$ and inward-pointing asters with scale $1/k_{0}$ and an aster size given by $L_p$, the Peclet length.
and (iii) {\it Aster lattice} - consisting of a square lattice of inward-pointing asters where the aster size is again $L_p$ and the `lattice spacing' is $1/k_0$ ($\geq L_p$).

To see why asters should settle into a square lattice, note that in the limit $\lambda=0$, the right side of (\ref{nequation}) can be written as a derivative of an `energy-functional', thus steady state solutions are minimisers of this `energy' -- we find that the `energy density' of asters (in units of $K_1$) arranged in a square ($E_{sq}= -35 $) is lower than in a triangular ($E_{tr} = -29 $)  unit cell. As a check, we have verified that this square lattice persists when we reduce the spatial discretization to be much smaller than the aster lattice spacing.

\begin{figure}[t]
\begin{centering}
\includegraphics[angle=0,scale=0.6]{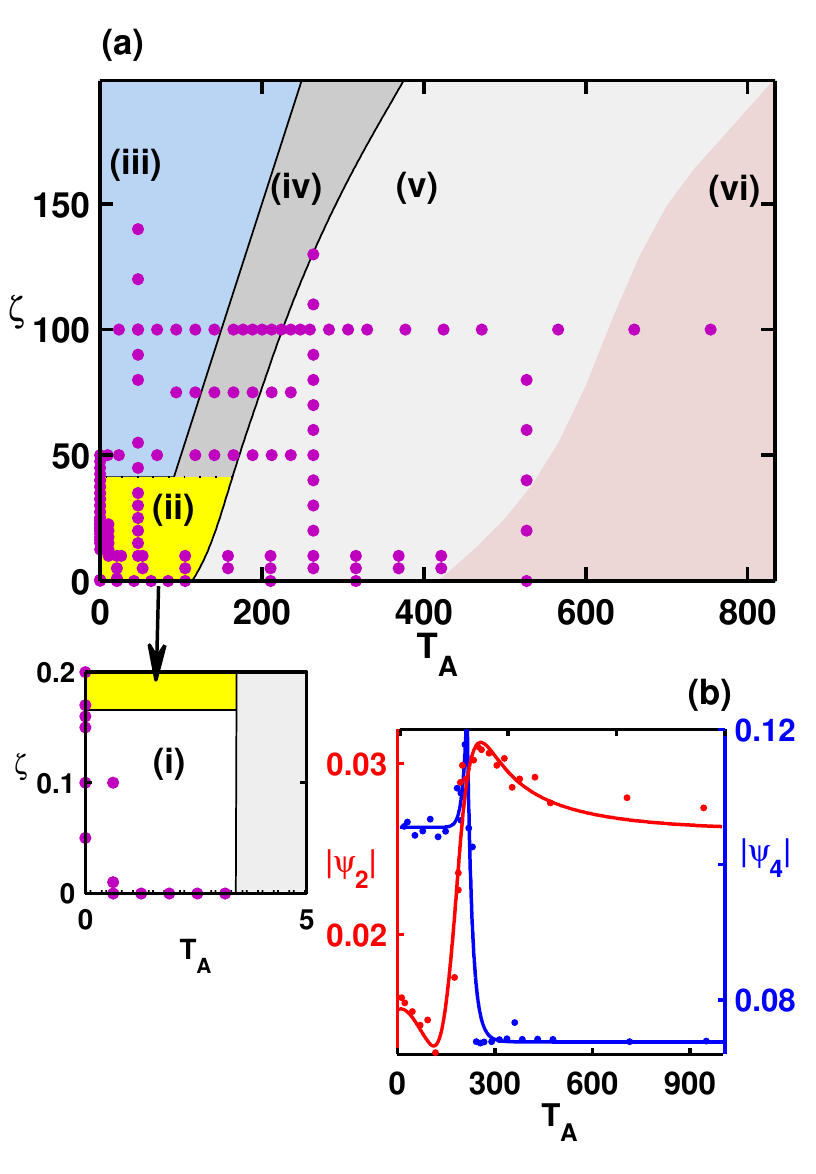}
\par\end{centering}
\caption{(a)  $\zeta$-$T_A$ phase diagram where the phases are (i) Boojum stripes (ii) Aster street (iii) Aster Lattice (iv) Bond tetratic (v) Bond Nematic and (vi) Isotropic phases. 
 (b) Variation of the bond orientational order parameters $\vert\Psi_2\vert$ ({\it nematic}, brown) and $\vert\Psi_4\vert$ ({\it tetratic}, blue) with $T_A$
 across the (iv) $\to$ (v) transition. Parameter values as in Fig.\,2.}
\end{figure}

Since $\langle {\bf n} \rangle \neq 0$ implies a movement of the active filaments with respect to the medium, the Boojum stripe and aster street are moving
phases, while the aster lattice is a stationary phase. The mean drift velocity given by 
$\vert\langle c{\bf n}\rangle\vert$ shows a  discontinuous jump across the Boojum-aster street transition 
and a continuous transition at the street-lattice phase boundary (Fig 1(B)), where we find $\vert\langle c{\bf n}\rangle\vert
\sim\vert\zeta-\zeta_{c}\vert^{\gamma}$ with $\gamma\approx1.38\pm0.05$. While we do not have an analytic explanation for the exponent value, it is easy to see why 
$\gamma > 1$ -- as one approaches the aster lattice phase from the street side, the filaments are drawn into the asters from the nearby parallel filaments by the $\zeta \nabla c$ term, as a result of which there is a nonlinear positive feedback which draws in filaments more strongly, enhancing the rate at which the net drift velocity vanishes.

Other defects such as vortices or outward-pointing asters are unstable, since the current ${\bf J} \propto c{\bf n}$ (\ref{cequation}). For instance, a vortex described by ${\bf n}
\propto {\hat {\bf e}}_{\theta}$ is unstable to radial fluctuations with a rate $\sim \zeta v_0/R$, where $R$ is the radius of the vortex. To stabilize such vortex configurations, one would need to include the binding to cross-linking proteins. One may obtain inward spiral asters, 
when the bend and splay distortions are comparable or even when the values of $\lambda$ are large. Being active such spiral asters would rotate with an angular velocity
\cite{curie1}.
The sequence of defect configurations (Boojum $\to$ aster $\to$ spiral) obtained here is similar to the sequence of achiral tilt textures in a circular  domain obtained on energy minimization \cite{pettey,sarasij}. In a different regime, 
 when the extrapolation length $L_e$ is large, one obtains transient configurations of moving walls with  filaments oriented normal to it. This resembles the configurations seen in
 reconstitution experiments \cite{bausch,shamim} and in recent numerical simulations \cite{shraddha}.

We now study the effects of active stochasticity on the steady state actin patterns by numerical integration of Eqs.\,(\ref{nequation}),(\ref{cequation}) with noise, whose strength is parametrized by the active temperature, $T_A >0$.  
At high $\zeta$, where the steady state is an aster (square) lattice at $T_A=0$, low noise results in phonon vibrations of the aster lattice. To quantify the state at low $T_A$, we 
compute the structure factor  $S({\bf q})$ from the fluctuations of the density of asters, which we 
define as $ \rho ({\bf r}) = -c({\bf r}) \nabla\cdot\bf{n}({\bf r})$. The structure factor shows  
 Bragg peaks indexed by the reciprocal lattice vectors of a square, the amplitude of the 
 peaks scales as $L^2$ for $0<T_A < T^*_A$ (Fig.\,2) - unlike
 the 2d equilibrium solid, the 2d active solid shows true long range order (LRO)
 at finite $T_A$ ! Beyond  $T^*_A$, the amplitude of the Bragg peaks scales as 
 $L^{2-\eta(T_A)}$, suggesting a transition to a solid with quasi-long range order (QLRO).
 We confirm this from a scaling plot of the structure factor $S(q)$ versus $qL^{2-\eta}$ 
 for $q\, \| \, {\bf G}$, the reciprocal lattice vector ${\bf G} = \left[1,1\right]$ (Fig. 2).
In addition we compute bond-orientational order parameters, namely 
 tetratic $\Psi_4 \equiv \langle e^{i4\theta}\rangle$ and 
nematic $\Psi_2 \equiv \langle e^{i2\theta}\rangle$, where $\theta$ is the orientation of the bonds between nearest neighbour asters and the ${\hat {\bf x}}$-axis. As seen in Fig.\,2,  the bond orientational order parameters $\Psi_4$ and
$\Psi_2$ clearly indicates a discontinuous transition from a QLRO solid 
with tetratic order to a bond-nematic liquid.
This jump in the order parameter $\Psi_2$ decreases on decreasing $\zeta$ and approaches zero at a multicritical point.

These structural transitions driven by the activity temperature $T_A$ are associated with a remodeling of the asters; beyond the active solid phase the asters break and 
reappear transiently,
 resulting in a decrease in the mean aster density with increasing $T_A$ \cite{kripa-cell}. The distribution of 
 lifetimes of the asters is a power-law in the tetratic phase ($P(\tau) \sim \tau^{-2.7}$) and exponential in the nematic phase \cite{kripa-cell}. 
 
Interestingly, the decrease in the number density of asters is accompanied by an increase in the 
net polarization $\langle c {\bf n}\rangle$ which jumps from being zero in the solid to a nonzero value in the 
tetratic and nematic phases (Fig.\,2). Thus the effect of  the active noise is to increase the polar order, weaning away filaments from the asters. Simultaneously, we 
 find that orientational fluctuations about the ordered direction are massless. 
 Figure 3 shows a power-law fit to the (unconnected) correlation function
 $C({\bf q}) \equiv \langle  {\bf n}({\bf q}, t) \cdot {\bf n}(-{\bf q}, t) \rangle$
  versus $q_{\perp}$ (wave-vector
 perpendicular to the ordering direction, taken to be along ${\hat {\bf x}}$; deviations
 occur both at low $q_{\perp}$ (corresponding to system size $L$) and high (corresponding to
 distance between asters, which decreases with increasing $\zeta$). For comparison, we plot
 the form expected when $\zeta \leq 0$; in this case, an exact RG calculation demonstrates that
$C({\bf q}) = q_{\perp}^{-6/5}$ in $d=2$ \cite{toner-tu}. The agreement with our numerics 
suggests that even for small, positive
values of $\zeta$, when the filaments are focussing,
the orientational fluctuations in the high noise regime are controlled by the Toner-Tu fixed point.
With a further increase in  $T_A$, the system eventually settles into an orientationally disordered phase.
 

\begin{figure}
\includegraphics[scale=0.8]{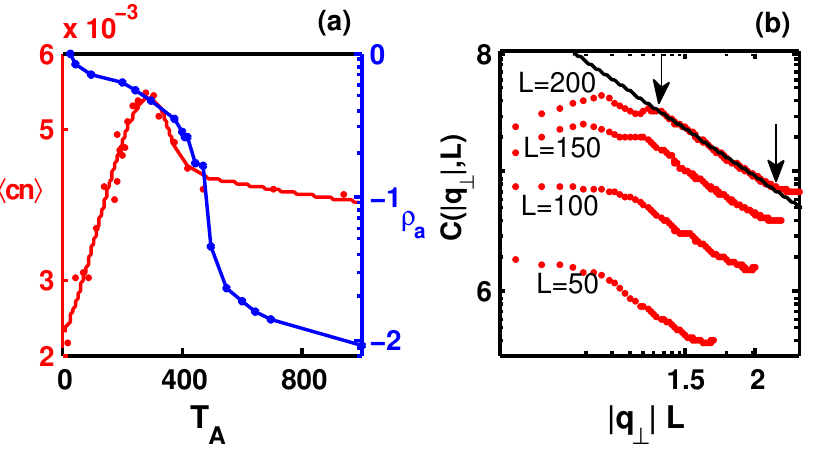}
\caption{(a) Sharp drop in density of filaments within asters (log plot in blue) with $T_A$ and the corresponding variation of the order parameter $c\bf{n}$ (red).
Parameter values same as Fig.\,2. (b) Logarithmic plot (base $10$) of the correlation function $C({\bf q}) \equiv \langle  {\bf n}({\bf q}, t) \cdot {\bf n}(-{\bf q}, t) \rangle$ vs $\vert q_{\perp}\vert L$ for the ordered (solid) and exponential phases (circles) for different system sizes. The black solid line shows fit to power law $C({\bf q}) \sim q^{-\theta}$
with $\theta=1.187\pm 0.061$, which compares well with the Toner-Tu exponent $\theta=6/5$.  Arrows show the lower (upper) wavevector cutoffs due to 
 system size (inter-aster distance).
Values of $\zeta= 10$ and $T_A=100$, rest same as Fig.\,1.}
\end{figure}
The sequence of transitions described above was in the high $\zeta$ regime. 
At lower values of $\zeta$, when the zero temperature phase is an aster street, an increase in $T_A$ drives the steady state directly into a bond nematic phase. 
Decreasing $\zeta$ further, i.e., starting from the Boojum stripe phase, we find that an increase in $T_A$ leads to elliptical domains of Boojums whose scale is set by  the correlation length, $L_c$ and $D/v_0$. In fact, even when $\zeta\leq0$ and small, a  nonzero $T_A$ produces these elliptical domains of Boojum, as a consequence of the advection current $v_0$.

Within  the calculation presented, the scale of the asters is set by the ratio $D/v_0$, though simple extensions of Eqs.\,(\ref{nequation}),(\ref{cequation}) to include higher order terms or other processes can lead to an increase in the 
 aster scale. One simple way is to increase the local concentration of filaments so as to attain a jammed 
aster configuration due to steric hinderance. Indeed the same result can be obtained with a lower concentration of filaments with the help of crosslinkers. An interesting alternative is to have a higher
depolymerization rate  $k_d$ at the aster cores balanced by a polymerization at the periphery. This would make the core size and hence the aster size bigger. A dramatic example of this phenomena is in the immunological synapse, where large actin asters, of order several microns, drives the clustering of T-cell receptors \cite{dustin}.

To conclude, we have made a detailed study of the nonequilibrium phases and transitions of active filaments in two dimensions, both with and without active noise.
The collective behaviour of active filaments gives rise to a variety of defect phases, which should be experimentally observable.
%
%
%
We thank S. Mayor, S. Ghosh, P. Srivastava, S. Ramaswamy and S. Sengupta for discussions and a critical reading of the manuscript.
 MR acknowledges research grants from HFSP and CEFIPRA-35104.


\begin{thebibliography}{44}

\bibitem{review} Chabbra E S and Higgs H N, \emph{Nat. Cell. Biol.}, \textbf{9}, 1110 - 1121 (2007). 

\bibitem{bausch} Schaller V, Weber C, Semmrich C, Frey E, Andreas R, Bausch A R, \emph{Nature}, \textbf{467}, 73-77(2010).

\bibitem{shamim} Butt T, Mufti T, Humayun A, Rosenthal P B, Khan S, Khan S and Molloy J, \emph{J. Biol. Chem}, \textbf{285}, 4964-4974 (2010).

\bibitem{sriramreview} Ramaswamy S, \emph{Ann. Rev. Cond. Matt. Phys.}, \textbf{1}, 323-345 (2010). 

\bibitem{joannyreview} Jülicher F, Kruse K, Prost J and Joanny J -F, \emph{Phys. Rep.}, \textbf{449}, 3-28 (2007). 

\bibitem{shraddha} Mishra S, Baskaran A and Marchetti M C, \emph{Phys. Rev. E}, \textbf{81}, 061916 (2010). 

\bibitem{simha-sriram} Simha R A, Ramaswamy S, \emph{Phys. Rev. Lett}, \textbf{89}, 058101 (2002).

\bibitem{curie1} Kruse K, Joanny J F, Jülicher F, Prost J and Sekimoto K, \emph{Phys. Rev. Lett.} \textbf{92}, 078101 (2004).

\bibitem{marchetti1} Liverpool T B and Marchetti M C, \emph{Phys. Rev. Lett.}, \textbf{90}, 138102 (2003).

\bibitem{hatwalne} Hatwalne Y, Ramaswamy S, Rao M and Simha R A, \emph{Phys.Rev.Lett}, \textbf{92}, 118101 (2004).

\bibitem{toner-tu}Toner J and Tu Y, \emph{Phys.Rev.Lett.} \textbf{75}, 4326 (1995); \emph{Phys. Rev. E}, \textbf{58}, 4828-4858 (1998).

\bibitem{note-convective}In principle there are three nonlinear convective terms $\bf{n}\cdot\nabla \bf{n}$, $\nabla(\vert\bf{n}\vert^2)$ and $\bf{n}\nabla\cdot\bf{n}$, of which only the first is relevant in $d=2$ (see \cite{toner-tu}).

%

\bibitem{numrecipe}Press W, Teukolsky S, Vetterling W and Flannery B, \emph{Numerical Recipes}, Cambridge University Press (1992).

\bibitem{leekardar}Lee H Y and Kardar M, \emph{Phys. Rev. E} \textbf{64}, 056113 (2001).

\bibitem{sumithra}Sankararaman S, Menon G and Sunil Kumar P B, \emph{Phys. Rev. E} \textbf{70}, 031905 (2004).

\bibitem{pettey} Pettey D and Lubensky T C, \emph{Phys. Rev. E}, \textbf{59}, 1834 (1999). 

\bibitem{sarasij} Sarasij R C and Rao M, \emph{Phys. Rev. Lett.}, \textbf{88}, 088101 (2002). 

\bibitem{chaikinlubensky} Chaikin P and Lubensky T C, \emph{Principles of Condensed Matter Physics}, Cambridge University Press (2000).

\bibitem{kripa-cell} Gowrishankar K, Ghosh S, Saha S, Rumamol C, Mayor S and Rao M, to be published in \emph{Cell} (2012).


\bibitem{dustin}Kaizuka Y, Douglass A D, Varma R, Dustin M L and Vale R, \emph{Proc. Nat. Acad. Sc}. \textbf{104}, 20296 (2007).




\end{thebibliography}
\end{document}